\documentclass[a4paper,12pt]{article}
\usepackage{epsfig,epsf,latexsym}
\begin{document}
% 
% ************************************************************
%                         FIGURES 
%
%\label{fig:diag}           diag_es3.eps*  [SAMIM.UA8.SE_PAPER]diag.kumac
%\label{fig:elastic}     elastic_es3.eps*  [SAMIM.UA8.SE_PAPER]factordiag.kumac
%\label{fig:zeus}           zeus_es3.eps*  [SAMIM.UA8.SE_PAPER]plotzeus.kumac
%\label{fig:h1}               h1_es3.eps*  [SAMIM.UA8.SE_PAPER]ploth1.kumac 
%\label{fig:apvseps}     apvseps_es3.eps*  [SAMIM.UA8.SE_PAPER]apvseps.kumac
%\label{fig:epsvsq2}     epsvsq2_es3.eps*  [SAMIM.UA8.SE_PAPER]ploteps.kumac
%\label{fig:ratio}         ratio_es3.eps*  [SAMIM.UA8.SE_PAPER]f2d2f.kumac
%
% 
% ************************************************************
\newcommand{\x}{\cdot}
\newcommand{\ra}{\rightarrow}
\newcommand{\pom}{\mbox{${\rm \cal P}$omeron}}
\newcommand{\flux}{\mbox{$F_{{\cal P}/p}(t, \xi)$}}
\newcommand{\fluxpap}{\mbox{$F_{{\cal P}/p}^{\bar{p}p}(t, \xi)$}}
\newcommand{\fluxgmp}{\mbox{$F_{{\cal P}/p}^{ep}(t, \xi)$}}
\newcommand{\ap}{\mbox{$\bar{p}$}}
\newcommand{\pap}{\mbox{$\bar{p} p$}}
\newcommand{\SPS}{\mbox{S\pap S}}
\newcommand{\xp}{\mbox{$x_{p}$}}
\newcommand{\sumet}{\mbox{$\Sigma E_t$}}
\newcommand{\mpr}{\mbox{${m_p}$}}
\newcommand{\mpi}{\mbox{${m_\pi}$}}
\newcommand{\rs}{\mbox{$\sqrt{s}$}}
\newcommand{\rsp}{\mbox{$\sqrt{s'}$}}
\newcommand{\rsps}{\mbox{$\sqrt{s} = 630 $ GeV}}
\newcommand{\lum}{\mbox{$\int {\cal L} {dt}$}}
\newcommand{\T}{\mbox{$t$}}
\newcommand{\abt}{\mbox{${|t|}$}}
\newcommand{\di}{\mbox{d}}
\newcommand{\HS}{\mbox{$xG(x)=6x(1-x)^1$}}
\newcommand{\sigdifjets}{\mbox{$\sigma_{sd}^{jets}$}}
\newcommand{\sigpomjets}{\mbox{$\sigma_{{\cal P}p}^{jets}$}}
\newcommand{\sigdiftot}{\mbox{$\sigma_{\bar{p} p}^{\rm tot \, diff}$}}
\newcommand{\sigpomtot}{\mbox{$\sigma_{p {\cal P}}^{\rm tot}$}}
\newcommand{\sigpompom}{\mbox{$\sigma_{{\cal P} {\cal P}}^{\rm tot}$}}
\newcommand{\sigpptot}{\mbox{$\sigma_{pp}^{\rm tot}$}}
\newcommand{\sigpomzero}{\mbox{$\sigma_{{\cal P}p}^o$}}
\newcommand{\dsig}{\mbox{${d^2 \sigma }\over{d \xi dt}$}}
\newcommand{\alamb}{\mbox{$\overline{\Lambda^{\circ}}$}}
\newcommand{\lamb}{\mbox{$\Lambda^{\circ}$}} 
\newcommand{\peetee}{\mbox{${ p_t}$}}
\newcommand{\PRET}{\mbox{\Proton-\sumet}}
%
%C new symbols for erhan-schlein II paper ------------------
%\newcommand{\xpom}       {\mbox{$x_{pom}$}}
\newcommand{\xpom}  {\mbox{$x_{I \! \! P}$}} 
\newcommand{\pmm}         {\mbox{${\cal P}$}}
\newcommand{\gm}         {\mbox{$\gamma^{*}$}}
\newcommand{\gmp}         {\mbox{$\gamma^{*} p$}}
\newcommand{\siggp}      {\mbox{$\sigma_{\gamma^{*} p}^{\rm tot}$}}
\newcommand{\siggpm}     {\mbox{$\sigma_{\gamma^{*} {\cal P}}^{\rm tot}$}}
\newcommand{\FtwoDtwo}       {\mbox{$F_2^{D(2)}$}}
\newcommand{\FtwoDthree}       {\mbox{$F_2^{D(3)}$}}
\newcommand{\xpomFtwo}     {\mbox{$\xi \FtwoDthree$}}
\newcommand{\qsq}        {\mbox{$Q^2$}}
\newcommand{\mx}         {\mbox{$M_X$}}
\newcommand{\mxsq}       {\mbox{$M_X^2$}}
\newcommand{\w}          {\mbox{$W$}}
\newcommand{\wsq}        {\mbox{$W^2$}}
\newcommand{\fluxint}    {\mbox{$f_{{\cal P}/p}(\xi)$}}
\newcommand{\eps}        {\mbox{$\epsilon$}}
\newcommand{\alf}        {\mbox{$\alpha '$}}
\begin{titlepage}
\vspace{4cm}
\begin{flushright}  
{8 December, 2003}
\end{flushright}
\vspace{4ex}
\begin{center}
\LARGE
{\bf {\boldmath A new $\gamma ^{*} p/\pap$ factorization test in}}\\
{\bf {\boldmath diffraction, valid below $\qsq \sim 6$~GeV$^2$}}\\
\normalsize
\vspace{9 ex}
Samim Erhan and Peter Schlein \\
\vspace{3.0mm}
University of California$^{*}$, Los Angeles, California 
90095, USA. 
\\
\end{center}
\vspace{11 ex}
\begin{abstract}
One of the key experimental issues in high energy hadron 
physics is the extent to which data from the diffractive interaction mechanism 
may be described by a factorized formula which is the product of a universal 
term describing the probability of finding a \pom\ in a proton 
(loosely referred  
to as the ``\pom\ flux-factor'') and a term describing the \pom 's interaction 
with the other incident proton.
In the present paper, after demonstrating that existing data 
on diffractive $\gm p$ and \pap\ interactions show that the \pom\ 
flux-factor is 
not universal, we present the results of a new test of factorization in 
these interactions which does not rely on universality of the flux-factor.
The test is satisfied to within $\sim 20\%$ for $1 < \qsq < 6$~GeV$^2$ and 
$\beta \leq 0.2$ in the $\gm p$ interactions, suggesting that the reasons for    
non-universality of the flux-factor have a limited effect on the factorization 
itself.
However, a clear breakdown of this test is observed at larger \qsq .
Kharzeev and Levin suggest that this can be attributed to the onset of 
QCD evolution effects in the \pom 's structure. 
The breakdown occurs in a \qsq\ region which agrees with their estimates of a 
small \pom\ size.

.

\end{abstract} 
\vspace{2 ex}
\begin{center}
submitted to European Physical Journal C\\
\end{center}
\vspace{15 ex}
\rule[.5ex]{16cm}{.02cm}
$^{*}$\ Supported by U.S. National Science Foundation Grant PHY-9986703\\
\end{titlepage}
\setlength{\oddsidemargin}{0 cm}
\setlength{\evensidemargin}{0 cm}
\setlength{\topmargin}{0.5 cm}
\setlength{\textheight}{22 cm}
\setlength{\textwidth}{16 cm}
\setcounter{totalnumber}{20}
\clearpage\mbox{}\clearpage
\pagestyle{plain}
\setcounter{page}{1}
\section{Introduction}
\label{sect:intro}
\indent

\begin{figure}[hbt]
\begin{center}
\mbox{\epsfig{file= 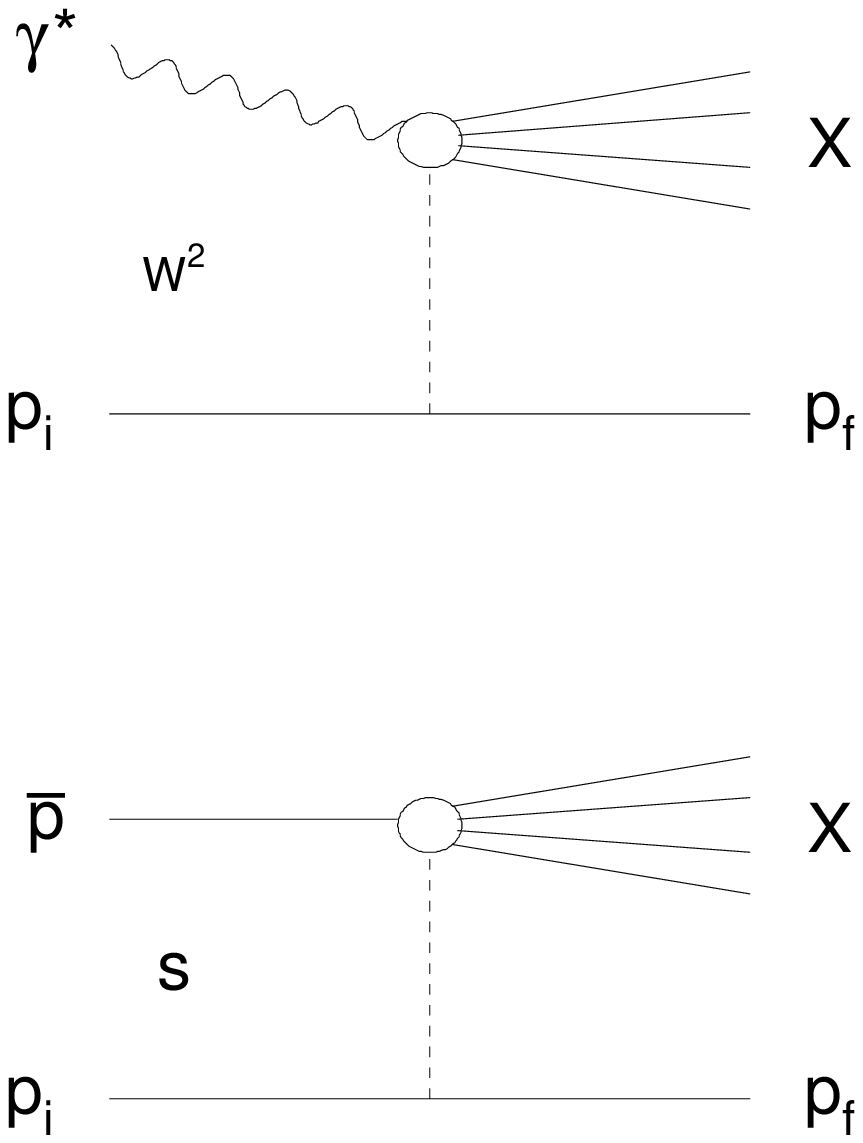,width=8cm}}
\end{center}
\caption[]{{\em
Upper: The diffractive $\gamma^{*}$-proton process. 
The total squared energy in the interaction is $W^2$. The 4-vector length 
squared of the $\gamma^{*}$
is -\qsq ; Lower: The diffractive \pap\ process. 
In each process, the exchanged \pom\ has a squared 4-momentum 
transfer, \T , and momentum fraction, $\xi \equiv \xpom = 1 - \xp$, of the 
incident proton.  
}}
\label{fig:diag}
\end{figure}

Studies of the inclusive inelastic production of beam-like 
particles
with momenta within a few percent of the associated incident 
beam 
momentum, as in:
\begin{equation}
\ap \, \,  \, \, + \, \, p_i \, \,  \, \, \ra \, \, \, \, X 
\, \, 
+ \, \, p_f 
\label{eq:dif}
\end{equation}
\begin{equation}
\gamma^{*} \, \, + \, \, p_i \, \,  \, \, \ra \, \, \, \, X 
\, \, 
+ \, \, p_f 
\label{eq:gams}
\end{equation}
have led to the development of a Regge 
phenomenology \cite{collins,basicphen}
of these processes (see Fig.~\ref{fig:diag}).
The observed final--state proton momentum
reflects the exchanged \pom 's momentum fraction
in the proton\footnote{We use the symbol $\xi$ for this 
variable in 
view
of its simplicity and its increasing use in the 
literature.}, 
$\xi \equiv x_{I \! \! P} = 1 - \xp$, and exchanged momentum 
transfer squared, 
\T . 

One of the relatively recent ideas~\cite{is} 
underlying the phenomenology is that, although
the \pom 's existence in the proton is due to non-perturbative
QCD, once the \pom\ exists, perturbative QCD processes can 
occur
in the proton-\pom\ and $\gamma^{*}$-\pom\ interactions
of Reacts.~\ref{eq:dif} and~\ref{eq:gams}, respectively. 
Ref.~\cite{is} proposed the study of such hard processes 
in Reacts.~\ref{eq:dif} and~\ref{eq:gams}
in order to determine the \pom 's structure.
Hard diffraction scattering was discovered in \pap\ 
interactions by
the UA8 experiment \cite{ua8} at the \SPS --Collider and in 
$ep$ 
interactions by the ZEUS \cite{zeusfirst} and H1 
\cite{h1first} 
experiments at HERA.

All available inclusive diffractive data from 
Reacts.~\ref{eq:dif} 
and~\ref{eq:gams} are well described 
\cite{ua8dif,zeus94,h194} 
by expressing the observed single-diffraction differential 
cross 
sections as 
products of factors describing the \pom\ flux in the proton, 
\flux\ 
(hereafter 
referred to loosely but conveniently as \pom\ emission), and 
\pom\ 
interaction,
for example proton--\pom\ or \gm --\pom\ total cross 
sections, 
respectively.
\begin{equation}
{{d^2 \sigma_{\bar{p} p}^{\rm diff}}\over{d \xi dt}} \, \, 
= \, \,  \fluxpap \, \, \x \, \, \sigpomtot(s')
\label{eq:factorhad}
\end{equation}
\begin{equation}
{{d^2 \sigma_{\gamma^{*} p}^{\rm diff}}\over{d \xi dt}} \, 
\, 
= \, \,  \fluxgmp \, \, \x \, \, \siggpm(s', \qsq )
\label{eq:factorgam}
\end{equation}
-\qsq\ is the squared momentum transfer
of the \gm\ in React.~\ref{eq:gams}.
$s'$ is the squared invariant mass of the $X$ systems in 
Reacts.~\ref{eq:dif} and~\ref{eq:gams}. 
To good approximation\footnote{The second equation comes 
from
$s' + \qsq - t = \xi (W^2 + \qsq)$, when $|t| << s'$ and 
$\xi << 
1$.},
$s' = \xi s$ in React.~\ref{eq:dif} and $s' = \xi W^2 -\qsq$ 
in
React.~\ref{eq:gams} (see Fig.~\ref{fig:diag}). 

There is, however, one complicating issue in the successful 
description
of the data by Eqs.~\ref{eq:factorhad} and 
\ref{eq:factorgam}.
The empirical \pom\ flux factors, \fluxpap\ and \fluxgmp , 
in the 
two equations 
are found to be different. 
More specifically, the effective \pom\ Regge trajectory in 
the 
common
factor, $\flux \sim \xi^{1-2\alpha (t)}$, required to fit 
the data
is different in React.~\ref{eq:dif} and \ref{eq:gams}.
The ZEUS \cite{zeus94} and H1 \cite{h194} collaborations 
both 
demonstrated
that the  effective \pom\ trajectory at low--$|t|$  in 
React.~\ref{eq:gams} lies above the effective trajectory 
which 
characterizes React.~\ref{eq:dif} (the evidence for this is 
shown 
below in 
Sect.~\ref{sect:gamphenom}). 
This is a remarkable situation and tells us that, although 
all 
existing data are 
well described by Eqs.~\ref{eq:factorhad} and 
\ref{eq:factorgam}, 
the \pom\ 
flux factor in the proton is not universal.

This conclusion should not come as a surprise because, 
for example, Kaidalov et al. \cite{kaidpomt} predicted that 
higher-order non-perturbative \pom -exchange effects in $pp$ interactions 
lead to an effective \pom\ Regge trajectory whose intercept at $t = 0$ 
decreases with increasing energy.
Moreover, as summarized in the following section, we have 
reported \cite{es2} such effects\footnote{Only below $s \sim 550$~GeV$^2$ is the 
effective \pom\ trajectory equal to the trajectory which describes the 
$s$-dependence of the $pp$ and $p\ap$ total cross sections \cite{cudell,dino2}.}
by fitting Eq.~\ref{eq:factorhad} to all available data on 
React.~\ref{eq:dif}.
Presumably, similar but weaker effects should also take place 
in $ep$ colisions. 

\begin{figure}[hbt]
\begin{center}
\mbox{\epsfig{file= 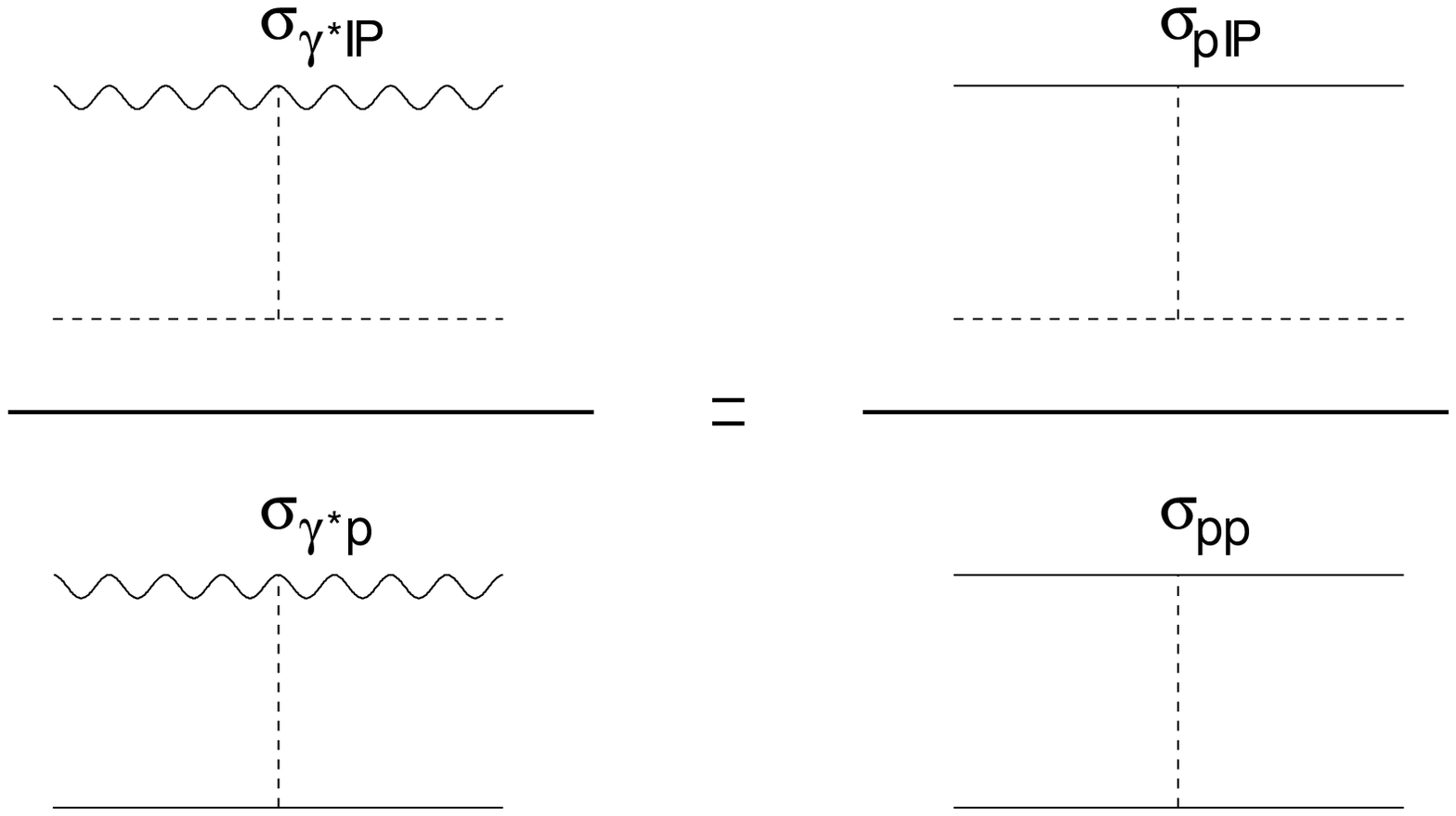,width=8cm}}
\end{center}
\caption[]{{\em
Ratios of the \gm\ and hadronic forward elastic amplitudes 
referred to in the text. 
In all cases, the dashed lines are \pom s, the solid lines are protons and the 
curved lines are $\gamma^*$s.
On both left and right sides, the upper vertices cancel, showing that
each is the ratio of \pom -\pom\ to \pom -proton vertices.
Hence, the left and right sides should be equal.
}}
\label{fig:elastic}
\end{figure}

In the present paper, despite the non-universality of the 
\pom\ flux factor in Reacts.~\ref{eq:dif} and \ref{eq:gams}, we propose to test 
the factorization represented in  Eqs.~\ref{eq:factorhad} and 
\ref{eq:factorgam}.
We accomplish this by asking if the \pom --exchange 
components of the extracted $\gamma ^*$-\pom\ and $p$-\pom\ total cross 
sections satisfy the relationship:
\begin{equation}
{{\siggpm}\over{\siggp}} \, \, = \, \, 
{{\sigpomtot}\over{\sigpptot}},
\label{eq:ratiosig}
\end{equation}
where the denominators are the total $\gamma^{*} p$ and 
$pp$ cross sections, respectively, and where all four are 
evaluated at the same cms interaction energy~\cite{ryskin}.
Equation~\ref{eq:ratiosig} is obtained from the optical 
theorem and the ratios of the forward elastic amplitudes shown in 
Fig.~\ref{fig:elastic}.

However, Eq.~\ref{eq:ratiosig} can not be used directly.
In extracting the $\gamma ^*$-\pom\ and $p$-\pom\ cross 
sections from the data using Eqs.~\ref{eq:factorhad} 
and~\ref{eq:factorgam}, in each case only the product of a flux factor 
normalization constant, $K$, and the cross section is experimentally accessible.
However, since \flux\ is not universal, $K$ may also not be 
universal. 
Thus, we introduce the notation, $K_{ep}$ and  $K_{pp}$ for 
the two cases, respectively, and the test of Eq.~\ref{eq:ratiosig} 
is actually a test of:
\begin{equation}
{{K_{ep} \, \siggpm}\over{\siggp}} \, \, = \, \, 
{{K_{pp} \, \sigpomtot}\over{\sigpptot}},
\label{eq:Kratio}
\end{equation}
If Eq.~\ref{eq:Kratio} is found to agree with data, as seems 
to be 
the case (see Sects.~\ref{sect:factor} and 
\ref{sect:conclude}),
the simplest explanation is that $K_{ep} \approx K_{pp}$ and 
the 
extracted cross sections obey Eq.~\ref{eq:ratiosig}.

In Sect.~\ref{sect:difphenom} we review the existing 
phenomenological analyses 
of React.~\ref{eq:dif} in terms of Eq.~\ref{eq:factorhad}
and extract the right-hand side of Eq.~\ref{eq:Kratio}.
In Sect.~\ref{sect:gamphenom} the HERA diffractive data
on React.~\ref{eq:gams} is reanalyzed and the left-hand side 
of Eq.~\ref{eq:Kratio} is extracted from the data.
In Sect.~\ref{sect:factor}, the factorization test is 
carried out using Eq.~\ref{eq:Kratio}. 
Our conclusions are given in Sect.~\ref{sect:conclude}.

\section{Review of diffractive {\boldmath $\ap p$} and 
{\boldmath $pp$} data analysis}
\label{sect:difphenom}
\indent

The UA8 collaboration \cite{ua8dif} fit 
Eq.~\ref{eq:factorhad} to 
the joint 
$\xi$-$t$ distributions of the available data on 
React.~\ref{eq:dif} 
at the \SPS\ ($\rs = 630$~GeV and $1.0 < |t| < 2.0$~GeV$^2$) 
and the corresponding $pp$ data at the ISR~\cite{albrowisr} 
($\rs = 23.5, 30.5$~GeV and $|t| < 2.0$~GeV$^2$),
all with $0.03 < \xi < 0.09$.
They obtained parametrizations of \flux\ and \sigpomtot\ 
which embody some features not previously  known:
 
\begin{equation}
\flux = K_{pp} \cdot |F_1(t)|^2 \cdot e^{(1.1 \pm 0.2)t} 
\cdot \xi^{1-
2\alpha (t)} 
\label{eq:flux}
\end{equation}

\begin{equation}
\alpha(t) = 1 + \epsilon + \alpha ' t + \alpha '' t^2 = 
1.10 + 0.25 t + (0.079 \pm 0.012) t^2
\label{eq:traj}
\end{equation}

\begin{equation}
K_{pp} \, \sigpomtot (s') \, \, = 
\, \, (0.72 \pm 0.10) \cdot [(s')^{0.10} + 
(4.0 \pm 0.6) (s')^{-0.32}]\,\,\,\,\, {\rm mb \, GeV^{-2}}. 
\label{eq:spomprot}
\end{equation}
With  $|F_1(t)|^2$ in Eq.~\ref{eq:flux} set equal to the 
Donnachie-Landshoff \cite{dl1} form 
factor\footnote{$F_1(t)={{4 m_p^2 - 2.8t} \over{4 m_p ^2 - t}}\, \x \, 
{1\over{(1-t/0.71)^2}}$}, 
the additional exponential factor is required.

The fits show that the effective \pom\ Regge trajectory flattens in the domain, 
$1.0 < |t| < 2.0$~GeV$^2$, as described by
by the quadratic term in Eq.~\ref{eq:traj}, when $\epsilon$ 
and $\alpha '$ are fixed at 0.10 and 0.25 GeV$^{-2}$, respectively.
In Eq.~\ref{eq:spomprot}, with the 
exponents\footnote{In this formula and others like it, 
``$s'$" stands for ``$s'/s_0"$, where $s_0$ = 1~GeV$^2$.} 
of $s' = \xi s$ fixed at 0.10 and -0.32, respectively,
$K_{pp} \sigpomtot (s')$ requires the presence of  
both \pom -- and Reggeon--exchange terms, as shown. 
Ref.~\cite{ua8dif} confirms the flattening of the effective 
trajectory at larger $|t|$ values, as well as the presence 
of the Reggeon--exchange term in Eq.~\ref{eq:spomprot}, by fitting 
the observed $\xi$--dependences at fixed $t$ values 
when\footnote{The sensitivity of fitting at small $\xi$ 
comes from the fact that, at small momentum transfer, the rapid increase of 
$\xi^{1-2\alpha (0)} \sim 1/\xi^{1+2\epsilon}$ dominates the 
relatively weak dependence of \sigpomtot\ on $\xi$ (via $s' = \xi s$).} 
$\xi < 0.03$.
In the fits, the experimental resolution and geometrical 
acceptance are taken into account.

A description of the phenomenology of React.~\ref{eq:dif} is 
incomplete without inclusion of the explicit effects of multi--\pom --exchange.
It has been widely known for some time that the observed 
$s$--dependence
of the total single--diffractive cross section, \sigdiftot , 
is not described by Eq.~\ref{eq:factorhad} 
(integrated over $t$ and $\xi < 0.05$) with a fixed
\pom\ Regge trajectory. 
Such a calculated \sigdiftot\ rises rapidly with energy and soon violates 
unitarity, while the observed \sigdiftot\ tends to level off or plateau at high 
energy \cite{dino,es1}. 
Since there is no built-in mechanism in the single-\pom -exchange 
process of Fig.~\ref{fig:diag} to account for the plateauing 
of \sigdiftot , there have been continuing theoretical efforts to satisfy 
$s$--channel unitarity \cite{unitarity};
this effect is attributed to multiple--\pom --exchange and 
is referred to variously in the literature as damping, screening, shadowing or 
absorption.
Kaidalov et al.\ \cite{kaidpomt} showed that multi--\pom --exchange 
diagrams lead to an effective \pom\ trajectory whose $t = 0$ 
intercept decreases with increasing energy.

In order to assess these effects quantitatively, 
Ref.~\cite{es2} performed fits of Eq.~\ref{eq:factorhad} 
integrated over $\xi < 0.05$ to the $d\sigma /dt$   
of all available ISR \cite{isrdsdt} and \SPS\ 
\cite{ua8dif,ua4dsdt} data.
In fitting to the complete set of ISR $d\sigma/dt$ data
over the energy range, $s$ = 549 to 3840 GeV$^2$, the only 
free parameters in Eqs.~\ref{eq:flux}, \ref{eq:traj} and \ref{eq:spomprot} 
were those in the effective \pom\ trajectory, each of which was assumed 
to have a simple $s$--dependence. 
The fit results from Ref.~\cite{es2} are\footnote{The logarithms are to base 
10.}:
\begin{tabbing}
\hspace{3cm}\=$\epsilon  (s)$ \hspace{6mm}\= =\hspace{4mm}\=
$(0.096 \pm 0.004) - (0.019 \pm 0.005) \cdot \log (s/549)$.\\ 
\>$\alpha '  (s)$ \> = \> $(0.215 \pm 0.011) - (0.031 \pm 
0.012) \cdot \log (s/549)$.\\ 
\>$\alpha '' (s)$ \> = \> $(0.064 \pm 0.006) - (0.010 \pm 
0.006) \cdot \log (s/549)$.\\
\end{tabbing}
At the lowest ISR energy, $s = 549$~GeV$^2$,  $\epsilon = 
0.096$, 
$\alpha ' = 0.215$~GeV$^{-2}$ and $\alpha '' = 
0.064$~GeV$^{-4}$, 
while each
of these is seen to decrease with increasing energy.
This is consistent with fixing $\epsilon = 0.10$ and $\alpha 
' = 
0.25$ 
in the fits of Ref.~\cite{ua8dif}, since
the only low--$|t|$ data in those fits were at the lowest 
ISR 
energies.

Ref.~\cite{es2} finds that the effective \pom\ trajectory 
continues
to decrease at higher energy. 
At the \SPS , (\rs\ = 630~GeV), the effective trajectory is:

\begin{center}
$\alpha(t) \, = \, 1 + \epsilon + \alpha' t + \alpha'' t^2 
\, = \, 
1.035 + 0.165 t + 0.059 t^2$
\end{center}
Ref.~\cite{es2} also shows that this $\alpha (t)$ form is consistent with the 
published function \cite{cdf} that is said to describe the CDF data on 
React.~\ref{eq:dif} at the Tevatron.

For completeness, we note that the fits of 
Ref.~\cite{ua8dif} are in kinematic regions where 
multi-\pom -exchange effects in 
React.~\ref{eq:dif} seem to be smallest.
In Refs.~\cite{es2,es1} it is shown that the effective 
$\alpha (t)$ is relatively independent of $s$ at the low end 
of the ISR energy range and that there is no evidence for 
$s$--dependence of the effective $\alpha (t)$ in the 
$|t|$--range, 1-2 GeV$^2$ (its average value is $0.92 \pm 0.03$.).
Thus, multi--\pom --exchange effects appear to be mainly in 
the low--$\xi$, low--$|t|$ region \cite{es2,es1}, where most 
of the cross section is.

To prepare for the factorization test of 
Eq.~\ref{eq:Kratio}, we 
need to
evaluate its right--hand--side.
Its numerator is given by Eq.~\ref{eq:spomprot},
while its denominator is the $pp$ total cross section,
which we take from the fits of Refs.~\cite{cudell,dino2}:
\begin{equation}
\sigpptot =   18 \, s^{0.10} - 27 \, s^{-0.50} 
+ 55 \, s^{-0.32}\,\,\,\,\,{\rm mb}.
\label{eq:Ksigpp}
\end{equation}
Since we are interested only in the \pom\ exchange terms in
Eqs.~\ref{eq:spomprot} and~\ref{eq:Ksigpp}, we drop the 
Reggeon-exchange
terms in both numerator and denominator.
The right-hand-side of Eq.~\ref{eq:Kratio} is then given by:
\begin{equation}
{{K_{pp} \, \sigpomtot}\over{\sigpptot}} 
\, \, \, \, = \, \, \, \, 0.041 \pm 0.007 \, \, {\rm GeV}^{-
2}.  
\label{eq:ratioa}
\end{equation}

\section{Analysis of diffractive {\boldmath $\gamma ^{*}p$} 
data}
\label{sect:gamphenom}
\indent

\begin{figure}[hbt]
\begin{center}
\mbox{\epsfig{file= 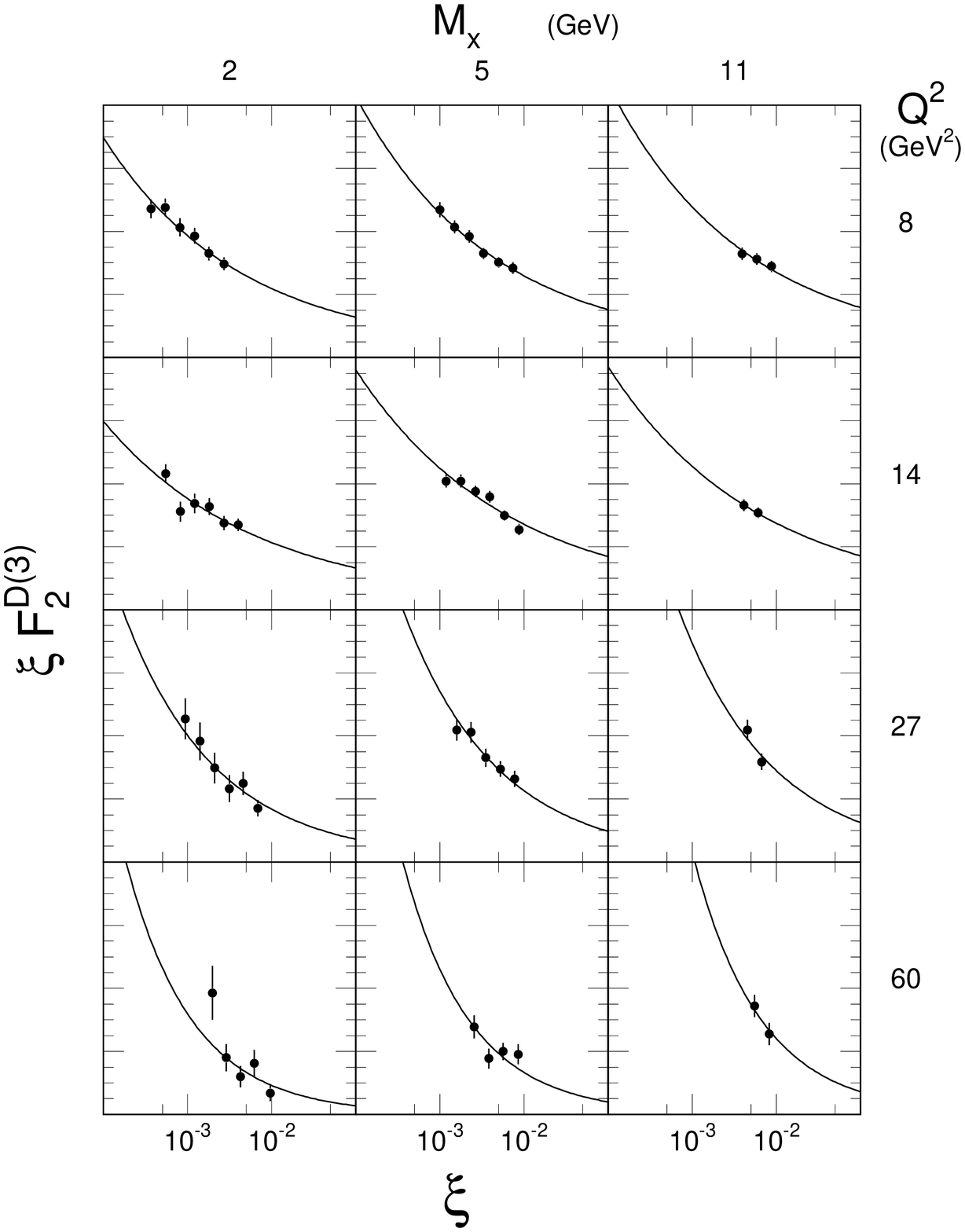,width=8cm}}
\end{center}
\caption[]{{\em
The ZEUS 1994 data\protect ~\cite{zeus94}: $\xi \FtwoDthree  $, 
vs.\ $ \xi$ ($\xi \equiv \xpom$) for 12 sets of (\mx , \qsq ) 
values.
At fixed \mx\ and \qsq , $\xi$ and \wsq\ are uniquely 
related ($\xi = (\mxsq + \qsq )/\wsq $). Thus, each set of 
points displays the \wsq\ dependence of React.~\ref{eq:gams} at 
fixed \mx\ and \qsq .  
The curves are the results of fitting Eq.~\protect\ref{eq:factorize2}
to the points shown, as discussed in the text.
}}
\label{fig:zeus}
\end{figure}

\begin{figure}[hbt]
\begin{center}
\mbox{\epsfig{file= 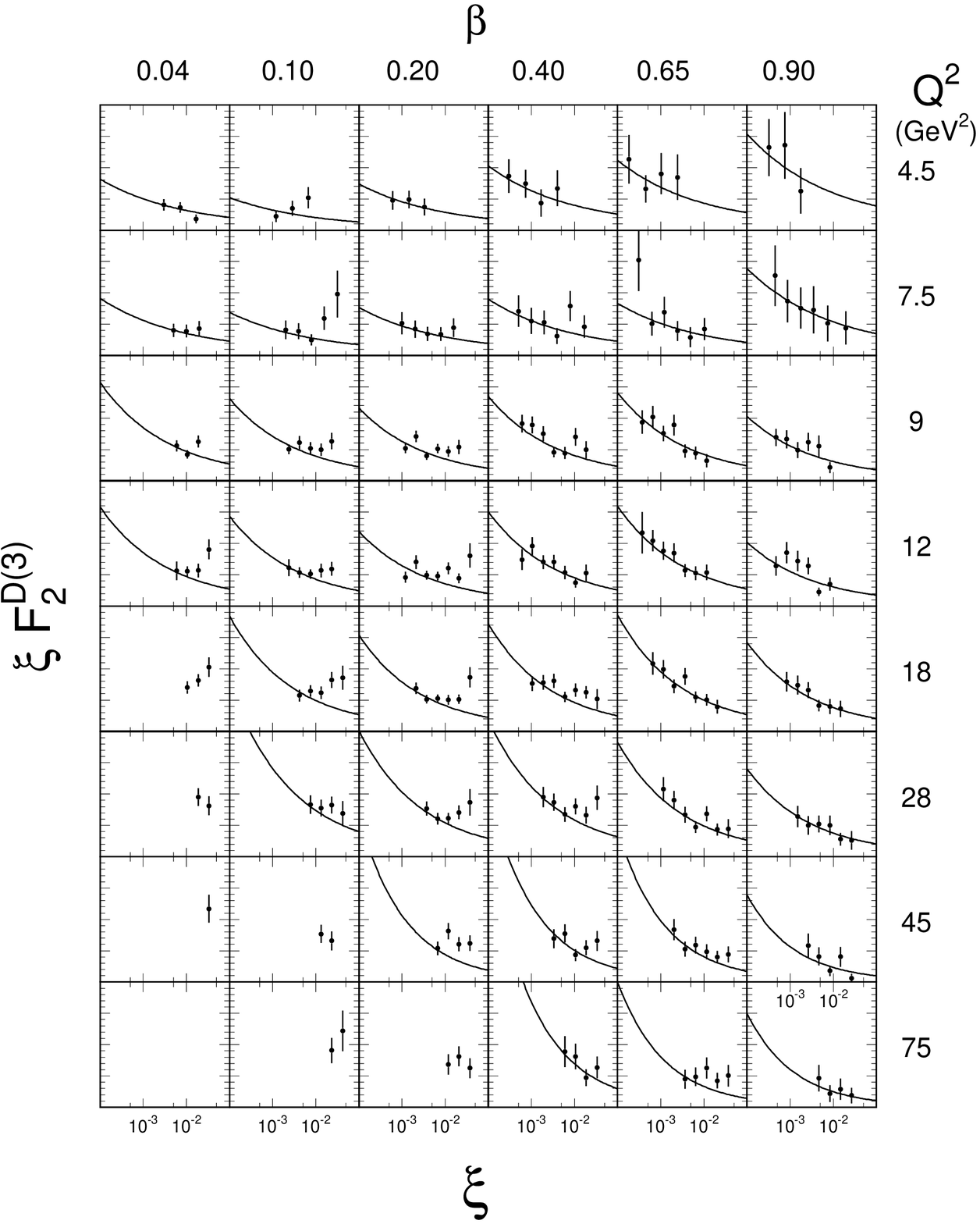,width=8cm}}
\end{center}
\caption[]{{\em
The H1 1994 data\protect ~\cite{h194}: $\xi \FtwoDthree  $, 
vs.\ $ \xi$ ($\xi \equiv \xpom$) in bins of $\beta$ and \qsq .
At each \qsq , $s' = \qsq \cdot (1-\beta)/\beta$.
The curves are the results of fitting Eq.~\protect\ref{eq:factorize2} to the 
points with $\xi < 10^{-2}$, as discussed in the text.
}}
\label{fig:h1}
\end{figure}

In order to carry out the factorization tests, we first 
reanalyze
HERA $ep$ diffractive data samples. 
The diffractive structure function, $\xi \FtwoDthree$, for 
the 
ZEUS 1994 data~\cite{zeus94} is displayed in 
Fig.~\ref{fig:zeus} 
and for the H1 1994 data~\cite{h194} in Fig.~\ref{fig:h1}.
The errors shown are obtained by summing the squared statistical and systematic 
errors, respectively.  
In both experiments, the recoil proton was not detected and the 
data are therefore integrated over \T .
This also means that the proton recoil system includes a low-mass excitation 
component, which was measured at the CERN ISR \cite{r603} to be $(12.0 \pm 
2.5)$\%  of the recoil system. 
This leads to a small systematic upward shift in the \FtwoDthree\ points,
which is corrected for in our final Fig.~\ref{fig:ratio}.

Following standard usage \cite{ingelpry,zeus94,h194}, the diffractive structure 
function, \FtwoDthree , is related to the diffractive $\gamma^{*}$--proton 
differential cross section by\footnote{Eq.~\ref{eq:f2d3} is obtained from Eq.~8 
of Ref.~\cite{zeus94}.}:
\begin{equation}
{{d\sigma_{\gamma^{*}p}^{diff}}\over{d \xi}} \, \, = \, \,
{{4 \pi^2 \alpha}\over{\qsq }} \cdot \FtwoDthree (\beta, 
\qsq , \xi)
\label{eq:f2d3}
\end{equation}
where, as noted earlier, the symbol, $\xi \equiv \xpom$, is used.

As in Ref.~\cite{is}, \FtwoDthree\ is written in factorized 
form, as the product of a \pom\ flux factor (in this case,
integrated over $t$) and a \pom\ structure function, \FtwoDtwo :
\begin{equation}
\FtwoDthree(\beta, \qsq , \xi) =  
\int \fluxgmp dt \, \cdot \, \FtwoDtwo (\beta, \qsq ) \, \, 
\approx \, \, {{K_{ep}}\over{\xi ^{1 + 2\eps} \cdot (3.9 - 2 \alf \, ln \, 
\xi)}}
\cdot \FtwoDtwo (\beta, \qsq ).
\label{eq:factorize1}
\end{equation}
or:
\begin{equation}
\xi \FtwoDthree(\beta, \qsq , \xi) \, \, \,  = \, \, \, 
{{K_{ep} \, \FtwoDtwo (\beta, \qsq )}\over
   {\xi ^{2\eps} \cdot (3.9 - 2 \alf \, ln \, \xi)}}.
\label{eq:factorize2}
\end{equation}
This approximate form of the flux factor integrated over $t$ arises from 
assuming $e^{3.9 t}\xi^{1-2\alpha(t)}$ for the functional form of \flux .
3.9 is the value which makes the integral equal to that of 
the $|t|$--integral of the full flux factor in Eq.~\ref{eq:flux} when 
\alf\ = 0.25 is used\footnote{This constant decreases to 3.7 and 3.5, for 
\alf\ = 0.15 and 0.05, respectively.}.

Fig.~\ref{fig:zeus} shows the fits of Eq.~\ref{eq:factorize2} to the 
ZEUS data.
The free parameters are \eps\ and an independent $K \FtwoDtwo$ at each of the 
twelve \qsq\ and \mx\ combinations (\alf\ is fixed at +0.25 GeV$^{-2}$). 
These fits, and those made to the H1 data in Fig.~\ref{fig:h1},
confirm factorization of \pom\ production and interaction in the diffractive 
\gmp\ interactions. 
In the domain, $\xi < 0.01$, where Reggeon exchange can be 
ignored at all $\beta$, all the observed dependence on $\xi$ is described by the 
flux factor in Eq.~\ref{eq:factorize2}.
Although this factorization has been known for some time 
from the H1 and ZEUS experiments, it is perhaps not widely recognized how 
remarkable it is.

\begin{figure}[hbt]
\begin{center}
\mbox{\epsfig{file= 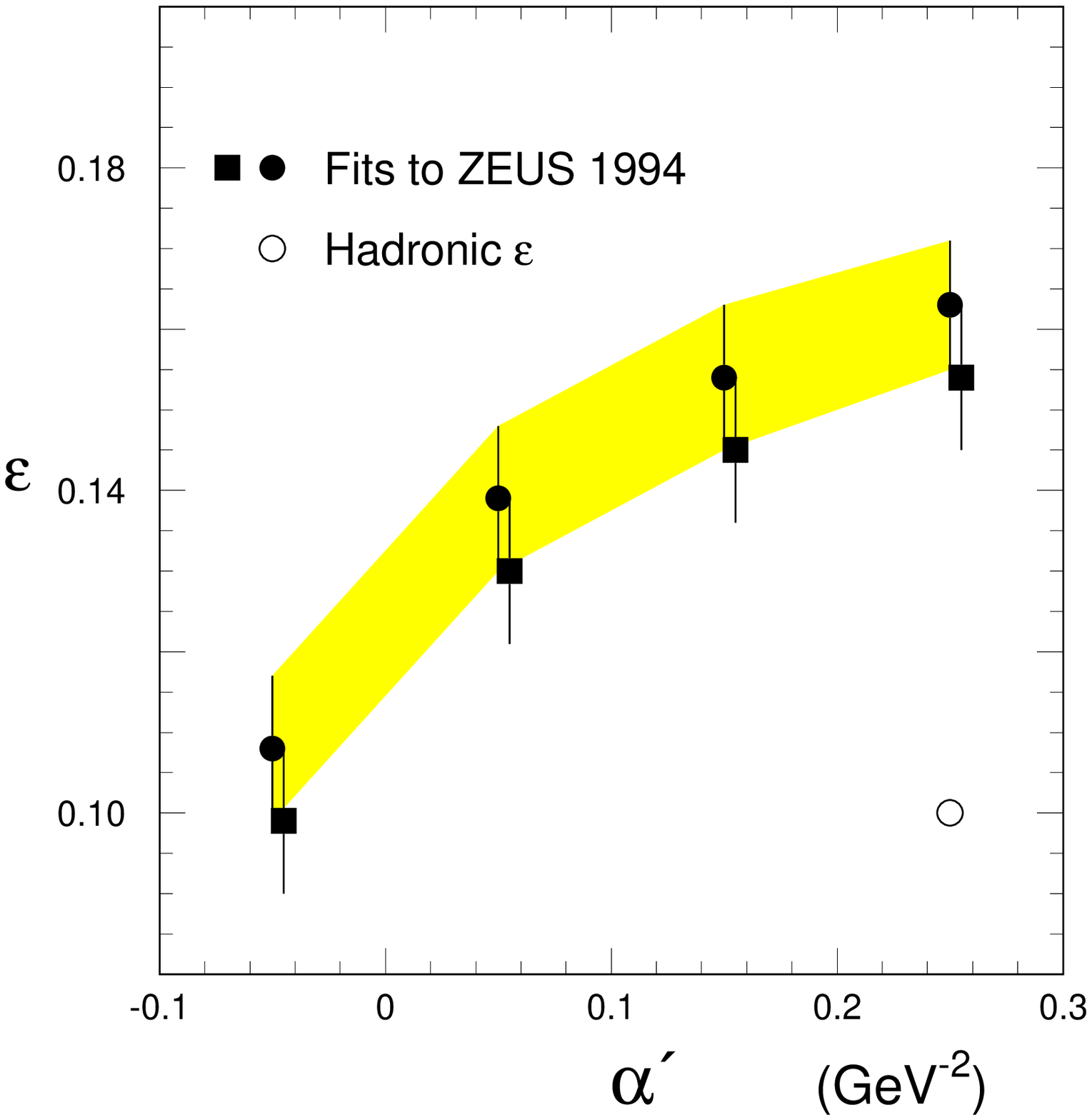,width=8cm}}
\end{center}
\caption[]{{\em
Fitted \eps\ vs.\ fixed $\alpha '$ from fits to the ZEUS 
1994 data shown in Fig.~\protect\ref{fig:zeus}. 
The solid circles are from fits to all four \qsq\ ZEUS data 
sets; the solid squares are from fits to only the two lowest \qsq\ 
data sets.
The shaded band represents the $\pm \sigma$ fit contour in the first set. 
The open circle shows the ``soft" \pom\ trajectory parameters, obtained 
from fitting the $s$--dependence of total $pp$ and \pap\ cross sections. 
}}
\label{fig:apvseps}
\end{figure}

From the fits to the ZEUS data, Fig.~\ref{fig:apvseps} shows 
the results when we fix \alf\ at the series of four values shown and determine 
\eps\ and the twelve normalization constants in Fig.~\ref{fig:zeus}. 
Fig.~\ref{fig:apvseps} shows a 1$\sigma$ error ``band" of allowed \alf\ and 
\eps\ values. 
All points along the valley of the contour are equally 
acceptable as solutions and there is insignificant discrimination between them 
with the present data. 
Although we assume $\alpha ' = +0.25$ for the factorization analysis
in this paper, we note that if the true effective $\alpha '$ were
as small as +0.15, the final ratios used in the factorization analysis 
only change by about 10\% and do not effect our conclusions.

We note in Fig.~\ref{fig:apvseps} that the band of allowed \eps\ and \alf\
values for the $\gamma^{*}$--p interactions is seen to be {\it inconsistent} 
with the conventional hadronic ``soft-\pom " effective Regge trajectory 
parameters \cite{cudell,dino2,dl1}, \eps\ = 0.10 and \alf\ = +0.25. 
This had been noted earlier by both ZEUS \cite{zeus94} and 
H1 \cite{h194}, but only for the assumed value, \alf\ = +0.25. 
We note here that the disagreement holds no matter what value of \alf\ is 
assumed.

\begin{figure}[hbt]
\begin{center}
\mbox{\epsfig{file= 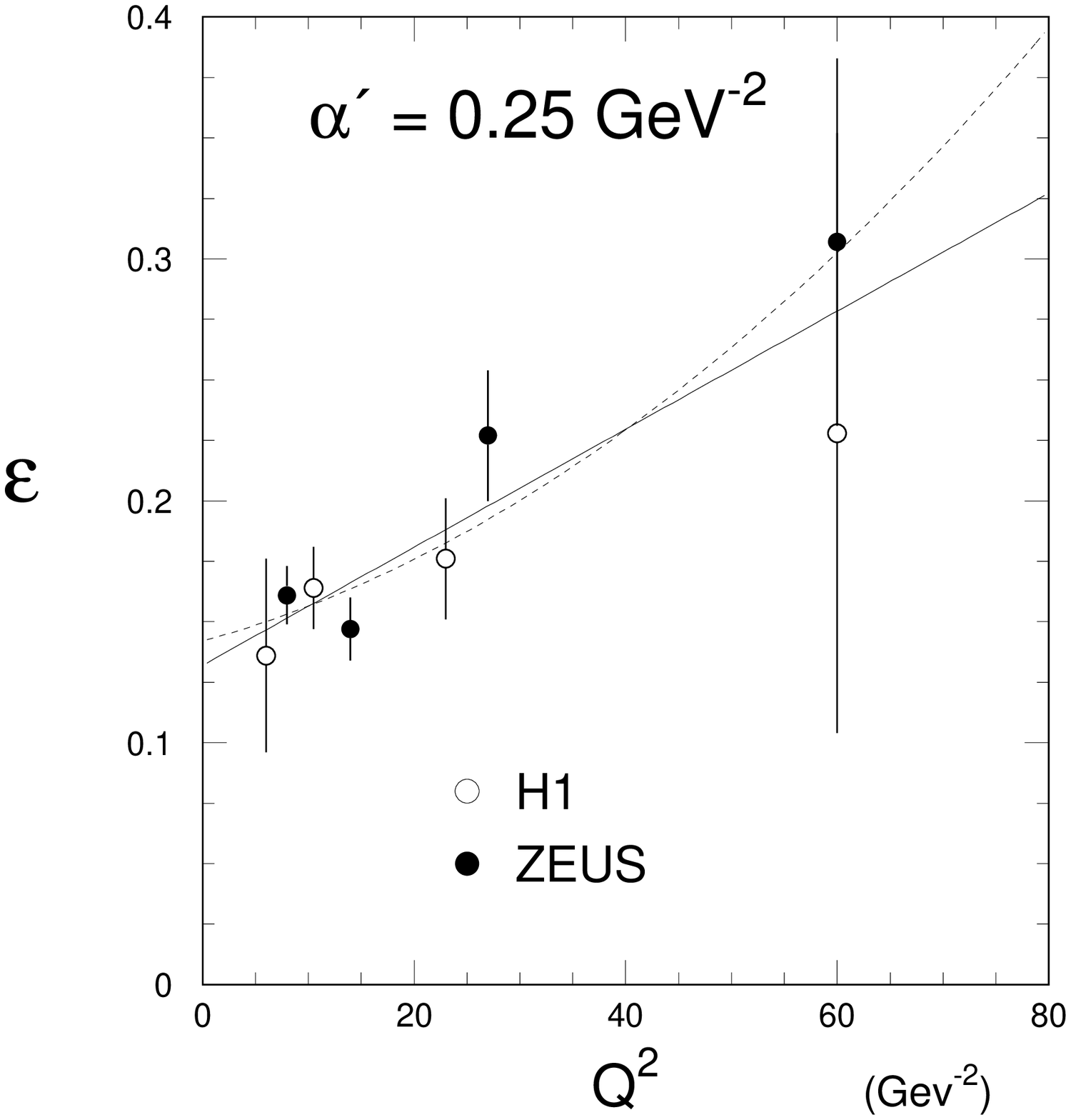,width=8cm}}
\end{center}
\caption[]{{\em
Fitted values of \eps\ vs.\ \qsq\ with \alf\ fixed at 
0.25~GeV$^{-2}$, as explained in the text. 
The solid and dashed curves are, respectively, linear and 
quadratic fits to the points shown.
}}
\label{fig:epsvsq2}
\end{figure}

Fig.~\ref{fig:epsvsq2} shows the fitted values of $\epsilon$ vs.\ \qsq\ for 
fixed \alf\ = +0.25. 
The ZEUS points correspond to simultaneous fits to the 
three distributions at each \qsq\ shown in Fig.~\ref{fig:zeus}. 
The H1 points are from combined fits made to the distributions with 
$\xi < 10^{-2}$, at each two neighboring \qsq\ values (4.5 and 7.5, 9 
and 12, etc.) shown in Fig.~\ref{fig:h1}.
There is a suggestion that \eps\ depends on $Q^2$.

\section{Factorization test}
\label{sect:factor}
\indent

To express the numerator on the left-hand-side of 
Eq.~\ref{eq:Kratio} 
in terms of the measured structure function, a comparison
of Eqs.~\ref{eq:f2d3} 
and \ref{eq:factorize1} with Eq.~\ref{eq:factorgam} yields:
\begin{equation}
K_{ep} \, \siggpm (s', \qsq ) \, \, = \, \, 
{{4 \pi^2 \alpha}\over{\qsq }} \cdot K_{ep} \, \FtwoDtwo 
(\beta, 
\qsq )
\label{eq:numer}
\end{equation}
where:
\begin{equation}
s' = \qsq (1-\beta)/\beta
\label{eq:sp}
\end{equation}

The denominator on the left-hand-side of Eq.~\ref{eq:Kratio}, \siggp , is 
approximately given in terms of the $F_2$ structure function by:
\begin{equation}
\siggp (W^2, \qsq ) \, \, = \, \, {{4 \pi^2 \alpha}\over{\qsq }} \cdot 
F_2 (x, \qsq)
\label{eq:denom}
\end{equation}
where, because both \siggpm\ and \siggp\ are evaluated at 
the same CM energy \cite{ryskin},  $W^2 = s'$ and
$x = \qsq /(W^2 + \qsq - m_p^2)$.
Our desired ratio is:
\begin{equation}
Ratio \, \, \equiv \, \, {{K_{ep} \, \siggpm}\over{\siggp}}  
\label{eq:ratiof}
\end{equation}

\begin{figure}[hbt]
\begin{center}
\mbox{\epsfig{file=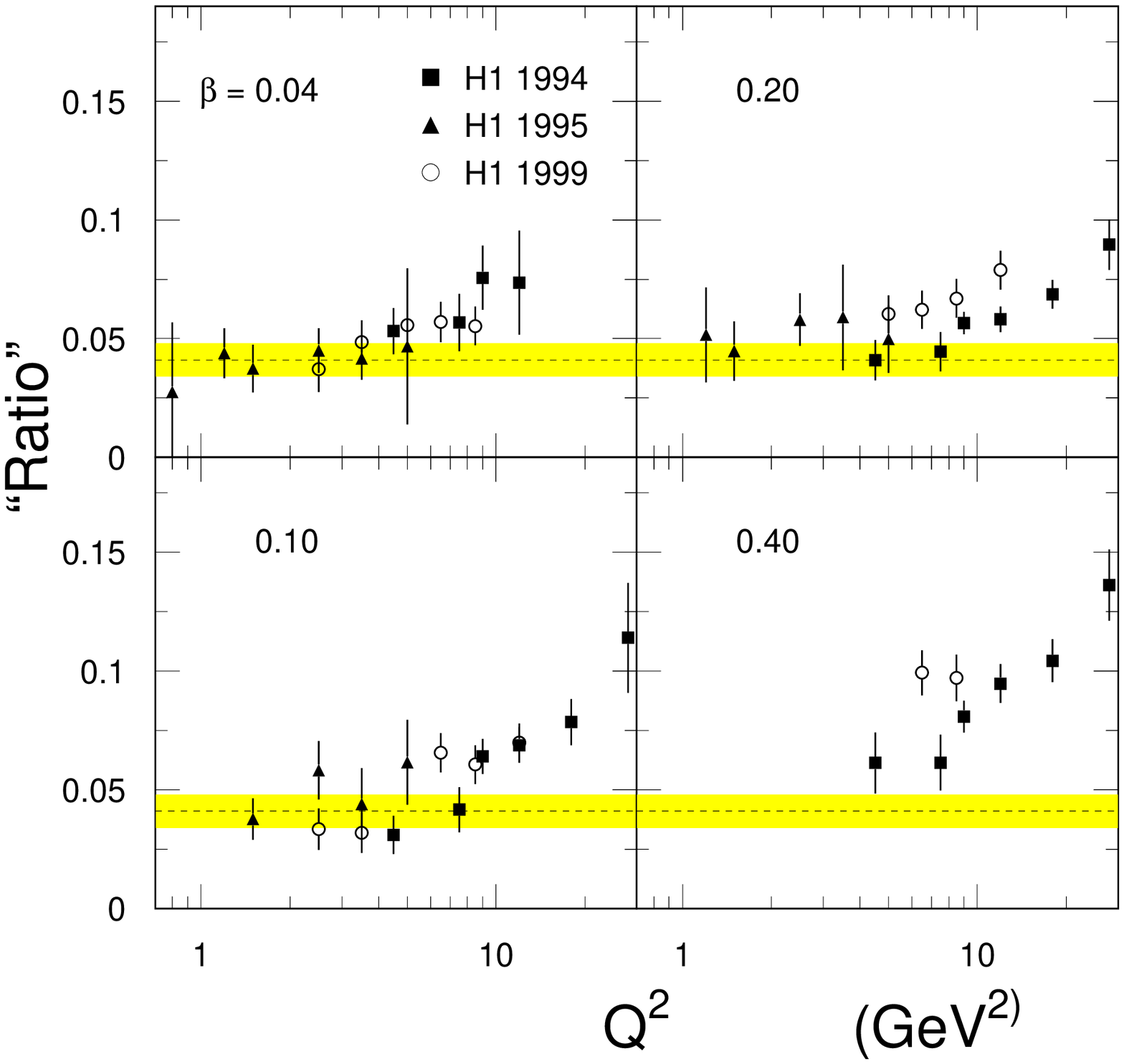,width=10cm}}
\end{center}
\caption[]{{\em
$Ratio$ defined in Eq.~\ref{eq:ratiof} vs.\ \qsq\ for 
$\beta$ = 0.04, 0.10, 0.20 and 0.4.
The solid-square points are calculated using the 1994 H1 
data \cite{h194} and are the ratios of $K_{ep} \, \siggpm$ to \siggp\ as 
explained in the text.
The solid triangles and open circle points use the 
preliminary lower--\qsq\ 1995 \cite{h1vanc} and 1999 \cite{h1amst} H1 data as 
measured from figures in their conference papers.  
The shaded band is the ratio of $K_{pp} \, \sigpomtot$ to 
\sigpptot ; see Eq.~\ref{eq:ratioa}.
In the extraction of the $K_{ep}\siggpm$ values, \eps\ = 0.15 and \alf\ = 0.25
are used. 
All points are systematically shifted downward by 12\% to account for excitation 
of the unobserved proton, as discussed in Sect.~\ref{sect:gamphenom}.
}}
\label{fig:ratio}
\end{figure}

Fig.~\ref{fig:ratio} shows the $Ratio$ defined in Eq.~\ref{eq:ratiof}
evaluated vs.\ \qsq\ for three different H1 data samples at 
$\beta$~=~0.04, 0.10, 0.20 and 0.40. 
For all points, the numberator is obtained from Eq.~\ref{eq:numer} and 
the denominator, \siggp , was calculated using the parameterization of 
Abramowicz and Levy \cite{allm}.

The solid-square points in Fig.~\ref{fig:ratio} with $4.5 < \qsq < 28$~GeV$^2$ 
are calculated using the 1994 H1 data \cite{h194}.
We see in the figure that these points are in reasonable 
agreement with the factorization prediction in Eq.~\ref{eq:ratioa}
at the lower end of their \qsq\ range (4.5 and 7.5 GeV$^2$) 
and $\beta < 0.4$,
while for \qsq\ larger than 6 or 7 GeV$^2$, there is a clear 
divergence from agreement with the prediction.
The points plotted in Fig.~\ref{fig:ratio} with solid 
triangles at lower \qsq\ values, $0.8 < \qsq < 5$~GeV$^2$, are preliminary 
results from the 1995 H1 data \cite{h1vanc}. 
Those plotted with open circles in the range, $2.5 < \qsq < 
12$~GeV$^2$, are preliminary results from the 1999 H1 data \cite{h1amst}. 
The points are seen to be in reasonable agreement in the \qsq\ domains where 
they overlap.
Eq.~\ref{eq:Kratio}, seems to be satisfied to within 20\% below $\qsq \sim 6$ or 
7~GeV$^2$ and $\beta < 0.4$.

As discussed in the following section, the breakdown of 
Factorization for \qsq\ above about 6 GeV$^2$ can be attributed to the 
onset of perturbative QCD efects on a small \pom . This is agreement with the 
magnitude calculated by Kharzeev and Levin \cite{kharzeev}.
However, several caveats concerning the results in 
Fig.~\ref{fig:ratio} should be noted.
\begin{enumerate}
\item The factorization prediction of Eq.~\ref{eq:Kratio}
is only valid for its \pom -exchange components. 
Although we have these for the right-hand-side of Eq.~\ref{eq:Kratio}, as shown 
in Eq.~\ref{eq:ratioa} and the bands in Fig.~\ref{fig:ratio}, we are presently 
unable to know the \pom -exchange components in the left-hand-side of 
Eq.~\ref{eq:Kratio}, or the data points in Fig.~\ref{fig:ratio}.
Thus, agreement between bands and data points in Fig.~\ref{fig:ratio}
implies that \pom -exchange is the same fraction of both numerator
and denominator of $Ratio$.
\item In calculating the points in Fig.~\ref{fig:ratio}, \eps\ = 0.15 and 
\alf\ = 0.25 are assumed. 
It is relevant to point out that the character of Fig.~\ref{fig:ratio} is not 
very sensitive to uncertainties in these parameters.
For example, as noted in the previous section, if $\alpha ' 
= +0.15$ is used to calculate the ratios in the figure, their values change by 
only $\sim 10\%$ and the conclusions do not change.
\item We pointed out above that the $K$ factor in \flux\ might be different in 
Reacts.~\ref{eq:dif} and \ref{eq:gams} and we therefore labeled them 
differently.
However, the approximate factorization agreement that we find at the 
lower \qsq\ values implies that the two $K$ values are probably not very 
different.
\end{enumerate}

\section{Conclusions}
\label{sect:conclude}
\indent

We have summarized the phenomenology of inclusive single 
diffraction in $pp$ ($\ap p$) interactions in which all available 
data are well described by a product of two functions which describe, 
respectively, the \pom\ flux in a proton and the \pom  --proton cross section. 
The \pom\ flux factor has the characteristic Regge form, except that the 
empirical \pom\ Regge trajectory is an effective one whose $t=0$ intercept and 
slope decrease with increasing energy. 
Following the arguments of Kaidalov et al. \cite{kaidpomt}, this presumably 
reflects multi--\pom --exchange processes which grow with energy, although it 
seems surprising that the factorized formula continues to describe the data as 
well as it does under these circumstances.

The data on diffractive $\gamma ^{*}p$ interactions in HERA 
$ep$ collisions are also well described by a product of a \pom\ flux factor 
and a cross section factor. 
However, the effective \pom\ trajectory is distinctly different from what is 
found in the corresponding hadronic interactions referred to in the previous 
paragraph.  
A possible interpretation for this fact is that multi--\pom --exchange effects 
are different in $pp$ and $ep$ collisions.  
 
From fits to the two diffractive data sets, we have extracted values for the
$\gamma^{*}$--\pom\  and $p$--\pom\ total cross sections (in each case 
multiplied by the normalization constant of the respective \pom\ flux factor).
We then combined these cross sections with the known total $\gamma^{*}p$ and 
$pp$ total cross sections to test a simple factorization relation between 
their \pom\ exchange components due to the optical theorem.  

The factorization test is observed to be reasonably well satisfied, to within 
about 20\%, in the range, $1 < \qsq < 6$~GeV$^2$. 
However, at higher \qsq\ values, a clear breakdown in the factorization test is 
observed.
 
In view of the pronounced and different \pom\ trajectory intercepts which are 
observed in the two classes of reactions, the first of these two observations is 
very surprising. 
It seems to be telling us that the dominant multi--\pom --exchange (or damping) 
effects are of such a nature that factorization of the \pom 's flux factor and 
its interaction is not very much disturbed.  

According to Kharzeev and Levin \cite{kharzeev}, our second observation 
concerning the breakdown of factorization observed at larger \qsq\ can be
understood in terms of the onset of QCD evolution effects in the \pom\ structure 
and a small \pom\ size (see also Ref.~\cite{kopeliovich}).  
Arguing that the properties of the soft \pom\ are linked to the scale anomaly of 
QCD, Kharzeev and Levin calculate that the scale, $M^2_0  \sim 4  \div 
6$~GeV$^2$ is the largest non--perturbative scale in QCD.
This corresponds to $\qsq \sim 1/R^2_{\cal P} \sim M^2_0$, 
where $R_{\cal P}$ is a typical size of the \pom .

The \qsq\ value at which our observed breakdown occurs gives 
a measure of the size of the \pom :  
$R^2_{\cal P} \sim 1/\qsq = (0.39$~GeV$^2$~mb)~/~(6~GeV$^2$) = 0.065~mb.
The area, $\pi R^2_{\cal P} = 0.20$~mb, agrees well with the recent UA8 
measurement \cite{pompom} of the \pom --\pom\ total cross section, 
\sigpompom\ = 0.2 mb above a center of mass energy of about 10 GeV. 
In this connection, it is also interesting to note that UA8 also obtained a 
statistically modest, but significant, test of factorization in 
double-\pom --exchange: 
 \begin{equation}
\bar{p} \, p  \,\, \rightarrow  \, \, \bar{p} \, X \, p
\label{eq:pompom}
\end{equation}
using the relation:
\begin{equation}
{{K^2 \, \sigpompom}\over{K \, \sigma^{tot}_{p{\cal P}}}} 
\, \, = \, \, {{K \, \sigma^{tot}_{p{\cal 
P}}}\over{\sigma^{tot}_{pp}}}
.\label{eq:sratio5}
\end{equation}
 
As pointed out in Sect.~\ref{sect:factor}, there are limitations to the
factorization analysis presented in this paper.
These can be addressed by obtaining larger and improved event samples.
For example, in the case of React.~\ref{eq:gams}, more detailed $t$-dependent 
measurements will allow an unambiguous determination of the effective \pom\ 
trajectory, espectially as a function of \qsq .
The left--hand side in Eq.~\ref{eq:Kratio} should be determined for the
\pom --exchange component alone. 
In the case of React.~\ref{eq:dif}, there is a great need for new and more 
complete data samples over a wider range of $t$ and $s$. 
This will be necessary in order to understand how the validity of our 
factorization test depends on the degree of damping in the reactions. 
This may be a very important issue, which we were not able to address in this 
paper because of a lack of the necessary data. 
To pursue this topic in the future, it will be necessary to have detailed 
studies of React.~\ref{eq:dif} at the Tevatron and a possible new experiment at
RHIC, which can cover the energy range between ISR and \SPS .

In summary, we conclude that, despite the non-universality of the \pom\ flux 
factor in Reacts.~\ref{eq:dif} and \ref{eq:gams}, the differential cross 
sections for these reactions can, to good approximation, still be written as a 
product of \pom\ formation and interaction factors.
In other words, there is no evidence for a breakdown of the factorization  
embodied in Eqs.~\ref{eq:factorhad} and \ref{eq:factorgam} for $\qsq < 
6$~GeV$^2$ in React.~\ref{eq:gams}.
The apparent factorization breakdowns reported in 
Refs.~\cite{whit,cdffact,covolan} are likely due to the different effective 
\pom\ trajectories in $ep$ and $pp$ interactions.
 
\section*{Acknowledgements}
\indent

We thank John Dainton and Guenter Wolf for extensive, 
invaluable 
discussions about diffractive $\gamma^{*}p$ interactions
and providing us with the 1994 H1 and ZEUS diffractive
data sets. 
We are greatly indebted to Dima Kharzeev for pointing out 
the 
significance 
of our results in terms of his work with Genya Levin.
We also thank 
Alexei Kaidalov, Boris Kopeliovich, Uri Maor and Misha 
Ryskin
for helpful discussions.
We are grateful to CERN, where most of this work was done,
for their continued hospitality.

\pagebreak


\begin{thebibliography}{99}

\bibitem{collins}
see: P.D.B.~Collins, ``An Introduction to Regge Theory \& 
High 
Energy Physics'',
Cambridge University Press (1977), and references therein.

\bibitem{basicphen}
See, for example:\\
A.H.~Mueller, Phys. Rev. D 2 (1970) 2963; D 4 (1971) 150;\\
A.B.~Kaidalov et al., Pisma JETP 17 (1973) 626;\\
A.~Capella, Phys. Rev. D 8 (1973) 2047;\\
R.D.~Field and G.C.~Fox, Nucl. Phys. B 80 (1974) 367;\\
D.P.~Roy and R.G.~Roberts, Nucl. Phys. B 77 (1974) 240;\\
A.B.~Kaidalov, Phys. Reports 50 (1979) 157.

\bibitem{is}
G. Ingelman and P.E. Schlein, Phys. Lett. B 152 (1985) 256.

\bibitem{ua8}
R. Bonino et al. (UA8 Collaboration), Phys. Lett. B 211 (1988) 239;\\
A. Brandt et al. (UA8 Collaboration), Phys. Lett. B 297 (1992) 417;\\
A. Brandt et al. (UA8 Collaboration), Phys. Lett. B 421 (1998) 395.

\bibitem{zeusfirst}
M. Derrick et al. (ZEUS Collaboration), Phys. Lett. B 315 (1993) 481.

\bibitem{h1first}
T. Ahmed et al. (H1 Collaboration), Nucl. Phys. B429 (1994) 477.

\bibitem{ua8dif}
A. Brandt et al. (UA8 Collaboration), Nucl. Phys. B 514 (1998) 3.

\bibitem{zeus94}
J. Breitweg et al. (ZEUS Collaboration), Europ. Phys. Jour. 
C 6 (1999) 43 and references therein.

\bibitem{h194}
C.~Adloff et al. (H1 Collaboration), Zeit. f. Physik C 76 (1997) 613.

\bibitem{kaidpomt}
A.B. Kaidalov, L.A. Ponomarov, K.A. Ter-Martirosyan,
Sov. Jour. Nucl. Phys. 44 (1986) 468.

\bibitem{es2}
S. Erhan and P. Schlein, Phys. Lett. B 481 (2000) 177.

\bibitem{cudell}
J.R.~Cudell, K.~Kang and S.K.~Kim, Phys. Lett. B 395 (1997) 
311; see also: ``Simple Model for Total Cross Sections",
preprint, Brown--HET--1060, January 1997.

\bibitem{dino2}
R.J.M.~Covolan, J.~Montanha \& K.~Goulianos, Phys. Lett. B 389 (1996) 176.

\bibitem{ryskin}
M.~Ryskin, private communication (1998).

\bibitem{albrowisr}
M.G. Albrow et al., Nucl. Phys. B54 (1973) 6;\\
M.G. Albrow et al., Nucl. Phys. B72 (1974) 376.

\bibitem{dl1}
A. Donnachie \& P.V. Landshoff, Nucl. Phys. B 231 (1984) 
189; Nucl. Phys. B267 (1986) 690; Nucl. Phys. B 244 (1984) 322.

\bibitem{dino}
K. Goulianos, Phys. Lett. B 358 (1995) 379; B 363 (1995) 268.

\bibitem{es1}
S. Erhan and P. Schlein, Phys. Lett. B 427 (1998) 389; B 445 (1999) 455.

\bibitem{unitarity}
See, for example:\\
A.~Capella, J.~Kaplan and J.~Tran Thanh Van, Nucl. Phys. B 105 (1976) 333;\\
V.A. Abramovsky, R.G. Betman, Sov. Jour. Nucl. Phys. 49 (1989) 747;\\
P.~Aurenche et al., Phys. Rev. D 45 (1992) 92;\\
E.~Gotsman, E.M.~Levin and U.~Maor, Zeit. Phys. C 57 (1993) 667; 
Phys. Rev. D 49 (1994) R4321; Phys. Lett B 353 (1995) 526.

\bibitem{isrdsdt}
M.G. Albrow et al., Nucl. Phys. B 108 (1976) 1;\\
J.C.M. Armitage et al., Nucl. Phys. B 194 (1982) 365.

\bibitem{ua4dsdt}
M. Bozzo et al. (UA4 Collaboration), Phys. Lett. B 136 (1984) 217;\\
D. Bernard et al., Phys. Lett. B 186 (1987) 227.

\bibitem{cdf}
F. Abe et al. (CDF Collaboration), Phys. Rev. 50 (1994) 5535.
 
\bibitem{r603}
L. Baksay et al. (R603 Collaboration) Phys. Lett. B 55 (1975) 491.

\bibitem{ingelpry}
G.~Ingelman and K.~Prytz, Zeit. f. Physik C 58 (1993) 285;
Proc. Workshop ``Physics at HERA'', ed. W. Buchmueller and
G. Ingelman, DESY 1992, Vol. 1, 233.

\bibitem{allm}
H.~Abramowicz and A.~Levy, ``The ALLM parametrization of $\sigma_{tot} (\gmp)$; 
an update'', DESY 97-251, hep-ph/9712415 (and private communication from the  
authors: There are misprints in Table 2; the quoted values of
$b_{{\cal P}1}, b_{{\cal P}2}, b_{{\cal R}1}, b_{{\cal R}2}$ should be squared.
In Eq. 3, both occurrences of $Q_o^2$ should be replaced by $Q_o^2 + \Lambda^2$.  

\bibitem{h1vanc}
H1 Collaboration, "Measurements of the diffractive structure 
function, \FtwoDthree , at low and high Q$^2$ at HERA", submitted to 
the 29th International Conference on High-Energy Physics ICHEP98, Vancouver, 
Canada (July 1998) Abstract 571. 
The \xpomFtwo\ values we use at $\xi = 0.005$ were obtained 
from their analog representions in Fig.~4.

\bibitem{h1amst}
H1 Collaboration, "Measurement of the Diffractive Deep-
Inelastic Scattering Cross Section at Low \qsq "
submitted to the 31st International Conference on High-Energy Physics 
ICHEP02, Amsterdam (July 2002) Abstract 981. 
The \xpomFtwo\ values we use at $\xi = 0.005$ were obtained 
from their analog representions in Fig.~6.

\bibitem{kharzeev}
D. Kharzeev and E. Levin, Nucl. Phys. B578 (2000) 351, hep-
ph/9912216, and private communications.

\bibitem{kopeliovich}
B.Z.~Kopeliovich, "From hard to soft diffraction and 
return", hep-ph/0012104.

\bibitem{pompom}
A.~Brandt et al. (UA8 Collaboration), Eur. Phys. Jour. C 25 (2002) 361.

\bibitem{whit}
L.~Alvero, J.C.~Collins, J.~Terron and J.J.~Whitmore,
Phys. Rev. D59 (1999) 074022.

\bibitem{cdffact}
T. Affolder et al. (CDF Collaboration), Phys. Rev. Lett. 84 (2000) 5043.

\bibitem{covolan}
R.J.M Covolan and M.S. Soares, Phys. Rev. D 60 (1999) 054005.


\end{thebibliography}
\end{document}